\documentclass[conference]{IEEEtran}
\usepackage{cite}
\usepackage{amssymb,amsfonts}
\usepackage{graphicx}
\usepackage{textcomp}
\usepackage{xcolor}
\usepackage{amsmath}
\usepackage{algorithm}
\usepackage[noend]{algpseudocode}
\usepackage{balance}

\usepackage{background}
\usepackage{xcolor}
\usepackage{hyperref}

\hypersetup{
  pdfborder={0 0 0}, 
}

\makeatletter
\algrenewcommand\alglinenumber[1]{}

\makeatother
\newcommand{\LineNo}[1]{\makebox[2.2em][l]{\footnotesize\textbf{#1}}}

\def\BibTeX{{\rm B\kern-.05em{\sc i\kern-.025em b}\kern-.08em
    T\kern-.1667em\lower.7ex\hbox{E}\kern-.125emX}}

\backgroundsetup{
  scale=1,
  color=black,
  opacity=1,
  angle=0,
  position=current page.south,
  vshift=10pt,
  contents={\textcolor{red}{© 2025. For personal use only. Published version under doi:  \href{https://doi.org/10.1109/MSWiM67937.2025.11309001}{10.1109/MSWiM67937.2025.11309001}. }}
}

\begin{document}

\title{Probabilistic Time Slot Leasing in TDMA-Based IoT Networks for Enhanced Channel Utilization}

\author{
\IEEEauthorblockN{Hicham Lakhlef}
\IEEEauthorblockA{Université de Bordeaux, CNRS\\
Bordeaux INP, LaBRI, UMR 5800\\
Talence, France\\
Email: hicham.lakhlef@labri.fr}
\and
\IEEEauthorblockN{Mohamed Ali Zormati}
\IEEEauthorblockA{Université de Technologie de Compiègne, CNRS\\ Heudiasyc, UMR 7253\\
Compiègne, France\\
Email: mohamed-ali.zormati@hds.utc.fr}
\and
\IEEEauthorblockN{Khaled Abid}
\IEEEauthorblockA{Université Paris-Saclay\\
CEA, LIST\\
Palaiseau, France\\
Email: khaled.abid@cea.fr}
\and
\IEEEauthorblockN{Toufik Ahmed}
\IEEEauthorblockA{Université de Bordeaux, CNRS\\
Bordeaux INP, LaBRI, UMR 5800\\
Talence, France\\
Email: tad@labri.fr}
}

\maketitle

\begin{abstract}
In large-scale resource-constrained wireless networks, such as those prevalent in the Internet of Things (IoT), efficient communication scheduling remains a critical challenge. Among the various approaches, Time Division Multiple Access (TDMA) protocols have been widely adopted for their structured and collision-free communication capabilities. Nevertheless, despite extensive research in this area, current solutions often exhibit suboptimal performance, particularly in dynamic environments where node activity levels fluctuate over time.

This paper introduces a novel fully distributed TDMA-based scheduling protocol that intelligently maximizes the utilization of communication resources. The proposed approach adaptively reallocates underutilized time slots, originally assigned to temporarily inactive nodes, to those experiencing higher communication demands. This dynamic reallocation not only improves channel utilization but also reduces idle periods, thereby enhancing overall network efficiency. To further enhance performance, we incorporate a lightweight probabilistic mechanism that governs the temporal leasing of unused slots. This mechanism balances the trade-off between slot availability and transmission reliability, minimizing packet loss while preserving fairness and stability within the network.

Simulations across a range of network scenarios demonstrate that our protocol significantly improves throughput, latency, and reliability in resource-constrained environments. These results highlight the protocol’s potential as a robust and scalable solution for adaptive and energy-efficient scheduling in next-generation IoT networks.
\end{abstract}

\begin{IEEEkeywords}
Internet of Things (IoT); constrained wireless networks; communication efficiency; time slot allocation.
\end{IEEEkeywords}

\section{Introduction}

The exponential growth of the Internet of Things (IoT) has led to billions of interconnected devices generating vast volumes of data across increasingly congested wireless networks \cite{zormatilakhlef2024}. These networks, often resource-constrained and bandwidth-limited, face significant challenges in maintaining reliable communication and delivering high Quality of Service (QoS) \cite{zikria2020deep, saleem2020intelligent}. Among these challenges, interference remains one of the most critical obstacles, severely affecting widely-used technologies such as Wi-Fi, Bluetooth, and ZigBee. Interference not only degrades QoS but also increases energy consumption through collisions and retransmissions \cite{saleem}.

To ensure robust and energy-efficient communication, it is essential to adopt strategies that minimize interference while maintaining timely and accurate data exchange, especially in large-scale deployments. One effective approach is to prevent message collisions, which are a primary source of interference in wireless environments. This leads us to consider scheduling techniques that can orchestrate transmissions while avoiding simultaneous access to the medium.

Among these techniques, Time Division Multiple Access (TDMA)-based protocols stand out for their deterministic behavior and low collision rates. Particularly in synchronous systems, which are common in IoT applications, TDMA-based schemes divide communication into structured rounds and slots. In each round, nodes execute a sequence of time slots where a message sent at the start of a slot is received by its destination at the end of that same slot \cite{mag20, R10}. However, when two neighboring nodes transmit in the same slot or a node receives multiple messages simultaneously, collisions occur, diminishing the system's reliability and efficiency.

To mitigate such collisions, a widely adopted method is distributed distance-2 coloring, in which each node is assigned a unique color (interpreted as a time slot), such that no two nodes within two hops share the same color \cite{lakhlefabid2024}. This technique has been applied in various distributed settings, including synchronous message-passing systems \cite{R10, BE14}, wireless sensor networks (WSNs) \cite{serena14, Ramanathan22}, shared memory models \cite{BGMBC22}, and self-stabilizing systems \cite{mag20}. These efforts aim to ensure that concurrent transmissions do not interfere with each other.

While these protocols effectively assign colors to avoid collisions, they typically assume a static or uniformly active network. In practice, however, node activity varies significantly over time. This results in a nonnegligible number of underutilized time slots, where inactive nodes still retain their assigned slots, effectively wasting communication opportunities and reducing overall network performance.

To address this limitation, we propose a novel distributed color-lending protocol that dynamically reclaims and reallocates these idle slots to more active neighbors. By allowing temporarily inactive nodes to lend their time slots to others, our protocol improves bandwidth utilization and reduces local buffer congestion, thus reducing energy consumption and improving QoS, especially in bursty or uneven traffic patterns.

This contribution fills an important gap in existing scheduling approaches by introducing adaptability and reusability into TDMA slot assignment, without introducing centralized coordination or significant overhead. The proposed solution is particularly well-suited for synchronous broadcast/receive systems, where communication precision and timing are critical.

The remainder of this paper is organized as follows. Section II reviews related work. Section III defines the system and communication models. Section IV presents our proposed underutilized time slot recovery protocol. Section V discusses the simulation results. Finally, Section VI concludes the paper and outlines directions for future research.

\section{Related Work}

In recent years, significant research efforts have been dedicated to improving resource utilization in constrained wireless networks. Several studies \cite{mag20, BE14, serena14, FLR16} have proposed distributed coloring algorithms to optimize time slot allocation and minimize interference in communication scheduling.

Despite their contributions, many of these protocols exhibit certain limitations. Some are not robust to collisions, while others rely on inefficient convergence procedures \cite{serena14}, or follow sequential communication approaches where only one node transmits at a time, which slows down the overall system performance \cite{FLR16}. Additionally, most of these methods do not consider multi-color assignments, where the same color (i.e., time slot) may be reused among potentially interfering nodes, an important consideration for dense and dynamic networks.

The work in \cite{Narasimha} explores a multichannel ultra-dense wireless network and proposes an age-based distributed Medium Access Control (MAC) mechanism, where each device adapts its idle channel probing rate based on message age. While adaptive, this approach does not fully address deterministic scheduling or explicit slot reuse.

In \cite{Lenka}, the authors propose a TDMA-based algorithm, Dynamic Slot Scheduling (DYSS), which aims to ensure collision-free, timely, and efficient communication in WSNs. It allocates time slots based on the average number of two-hop neighbors, rather than the maximum, as in \cite{Lenka1}. Although this reduces the overall number of slots in the schedule, it may leave nodes in densely populated areas without sufficient slot allocations. Compared to other approaches such as in \cite{Bhatia}, DYSS results in fewer total slots, reducing latency and improving collision handling.

Another relevant line of research involves $m$-frugal coloring \cite{molloy2010}, where adjacent nodes must receive different colors, and the same color may appear at most $m$ times in any node’s neighborhood. This model aims to reduce the probability of collisions in TDMA networks operating over $m$ channels. While the theoretical results are promising, showing that for large enough maximum degree $\Delta$, a $(\Delta+1)$-coloring exists, they are mostly asymptotic and do not translate easily to practical implementations in wireless settings.

Recent efforts have explored more adaptive and intelligent scheduling protocols tailored to the needs of dynamic IoT environments. For instance, Game-Theoretic Time-Slotted Channel Hopping (GT-TSCH) \cite{tavallaie2023gt} introduces a distributed scheduling strategy that adapts to traffic demands and link quality, thereby improving throughput and reducing delay in low-power wireless networks. Similarly, a Q-learning-based TDMA protocol for visible light IoT networks is proposed in \cite{makvandi2023machine}, enabling decentralized and collision-free slot assignment.

In \cite{dutta2023multi}, a multi-armed bandit approach is used to optimize TDMA slot selection under decentralized conditions, improving bandwidth utilization and adaptability to topology changes. Moreover, Proactive Reduction of Idle Listening Multi-hop with Latency constraints (PRIL-ML) \cite{scanzio2024wireless} and Balanced Space and Time-based (BST)-TDMA \cite{botirov2024balanced} demonstrate energy-aware and latency-sensitive scheduling solutions for industrial and light-based IoT networks, respectively.

While previous works, such as the distributed distance-2 coloring algorithm introduced in \cite{L3}, have laid the groundwork for secure and collision-free communication in general wireless environments, they often fall short in terms of adaptability and resource efficiency. These approaches typically assume static traffic patterns and uniform node activity, which is rarely the case in real-world IoT deployments characterized by bursty, heterogeneous, and dynamic communication demands.

Moreover, existing solutions lack mechanisms for the reuse or reallocation of underutilized time slots, leading to inefficient bandwidth usage and increased latency, particularly in dense or traffic-variable networks. Focusing on worst-case scenarios or asymptotic guarantees often limits practical performance in constrained environments.

In contrast, our work addresses these critical shortcomings by introducing a distributed, deterministic frugal coloring approach tailored to realistic and dynamic topologies. We incorporate a lightweight mechanism for reclaiming and reallocating unused slots, allowing nodes with higher communication demands to temporarily borrow slots from inactive peers. This not only improves channel utilization and reduces idle time but also maintains fairness and collision avoidance without requiring centralized control.

To the best of our knowledge, this level of adaptivity and slot-level flexibility has not been previously addressed in a fully distributed TDMA-based scheduling context, making our approach a promising candidate for scalable and energy-efficient communication in next-generation IoT systems.

\section{System and Communication Models}

In the following, we present the system model (i.e., nodes, initial knowledge, and communication graph), and the communication model.

\subsection{System model}
The system model consists of $n$ nodes, denoted as $p_i$, where $1 \leq i \leq n$, connected by an arbitrary communication graph $\mathcal{G}$ with maximum degree $\Delta$. Each node is capable of retrieving packets in two ways: either by sensing them from the environment or by receiving them from neighboring nodes. Note that the nodes themselves are not aware of their indices $i$, which is just a notation convenience.

Let $neighbors_i^{1,2,3}$ be a constant representing a local neighborhood set known by node $p_i$. It identifies its neighbors at distances 1, 2, and 3. Each node $p_i$ has its own unique identity $id_i$, which is known only to itself and the above neighbors. Specifically, $neighbors_i^{x}$, where $1\leq x\leq3$, denotes a set containing only the neighbors at distance $x$.

Each node in the network has a memory capacity, denoted $m_{size}(p_i)$, that allows it to store a finite number sensed packets at any given time. We recall that a node $p_i$ is responsible for sensing information that needs to be organized into packets and sent to a gateway for decision making. The sensing process of node $p_i$ follows a Poisson process with a rate of $\lambda_i$. Additionally, a node may receive information from other nodes that it must forward in order to deliver the packets to the gateway. Therefore, $m_{size}(p_i)$ is divided into two parts, $M$ and $F$, where $M$ is used to store sensed packets, while $F$ is used to store packets received from neighbors that need to be forwarded when it is $p_i$ turn to broadcast.

\subsection{Communication model}
In the network, communication is synchronous, which means that there is an upper bound $D$ on the delay of message transmission and this bound is known to all nodes. This knowledge is considered to be shared among all nodes. From an algorithmic point of view, we assume the existence of a global clock denoted $\mathit{Cl}$. The clock is incremented by 1 after each period of $D$ physical time units. Each value of $\mathit{Cl}$ corresponds to a specific time slot in the system.

Nodes can be in one of three states:
\begin{itemize}
    \item \textbf{Idle:} The node is neither transmitting nor receiving.
    \item \textbf{Transmitting:} The node is actively sending a packet.
    \item \textbf{Receiving:} The node is listening to the channel and can receive a packet, provided there is no interference.
\end{itemize}

In the considered network, a distributed algorithm similar to the one in \cite{L3} has already been used to perform a distance-2 coloring of the network. This coloring ensures that each node in the network is assigned a color from the set $\{0, ..., K\}$, where $K$ is less than or equal to $\Delta^2$. It has been proved that a graph of degree $\Delta$ can be distance-2 colored using at most $\Delta^2$ colors, as presented in \cite{Lloyd}.

The color assigned to a node $p_i$, noted as $color_i$, allows it to communicate without collisions. The time slots in which $p_i$ is allowed to send its messages are determined by the value of $\mathit{Cl}$ such that $\mathit{Cl}$ $\mod$ $\chi = color_i$. Here, we use $\chi = K+1$ to denote the number of colors (i.e., the number of available time slots).

Let $\mathit{Ncl^c_i}$ represent the maximal list $L$ composed of neighbors of node $p_i$, satisfying:

\begin{itemize} \item For all $x \in L$, $x$ does not have a node at distance 2 using color $c$, \item For all $x, y \in L$, $x \notin neighbors_y^{1,2}$ and $y \notin neighbors_x^{1,2}$. \end{itemize}
\subsection{Problem statement}

In a synchronous broadcast-based wireless network, each node is assigned specific transmission slots within a repeating frame. However, due to various factors such as changes in traffic patterns or node inactivity, some of these slots remain underutilized, leading to suboptimal use of the communication medium.

Our objective is to design a distributed protocol that enables nodes to locally detect and recover such underutilized transmission slots, while preserving the collision-free property of the schedule. More specifically, we aim to answer the following question:

\begin{quote}
\textit{How can a node opportunistically reuse underutilized slots assigned to other nodes, based solely on locally observable information, without introducing collisions?}
\end{quote}

The solution must satisfy the following constraints:

\begin{itemize}
    \item \textbf{Collision avoidance:} No two nodes within mutual interference range should transmit in the same slot.
    \item \textbf{Locality:} Decisions must be based only on 1-hop or 2-hop neighborhood information.
    \item \textbf{Timeliness:} Underutilized slots should be recovered promptly to maximize throughput.
\end{itemize}

The design space is constrained by the need to avoid global coordination while efficiently exploiting temporal and spatial diversity in slot usage. The proposed solution should adapt to changes in node behavior and traffic load with minimal overhead.

\section{Underutilized Time Slots Recovery Protocol}
In this section, we present our solution for recovering underutilized communication time slots in broadcast/receive synchronous wireless networks for constrained scenarios. The protocol is presented in Algorithm \ref{alg:utsr}.

\begin{algorithm*}
\caption{Underutilized time slots recovery algorithm}
\label{alg:utsr}

\noindent\textbf{Initialization:} $cnt_{0 \leq color \leq K} = 0;$ \quad $Candidat\_list_i = \emptyset;$ \quad $Bcolors_i = \emptyset;$

\vspace{0.3cm}
\noindent\textbf{Procedure \texttt{color\_lending()} for node $i$}

\begin{algorithmic}
\State \LineNo{01} \textbf{if} $Ncl^{color_i}_i \neq \emptyset$ \textbf{then}
\State \LineNo{02} $\mathcal{T}_i$\_$\mathcal{CC}$broadcast Lend\_color $((color_i, \mathcal{T}_{color_i}), Ncl^{color_i}_i)$;
\State \LineNo{03} \textbf{endif}
\State \LineNo{04} sleep $(\mathcal{T}_i)$;
\end{algorithmic}

\vspace{0.3cm}
\noindent\textbf{Procedure \texttt{adding\_lent\_color()} for node $i$}

\noindent\textbf{At the reception of Lend\_color $((color_j, \mathcal{T}_{color_j}), Ncl^{color_j}_j)$}

\begin{algorithmic}
\State \LineNo{05} \textbf{if} $(i \notin Ncl^{color_j}_j)$ \textbf{then} discard the message; \textbf{endif}
\State \LineNo{06} $\mathcal{T}_j$\_$\mathcal{CC}$broadcast interest\_in\_color $((color_j, \mathcal{T}_{color_j}), \lambda_i, id_i)$;
\end{algorithmic}

\vspace{0.2cm}
\noindent\textbf{At the reception of interest\_in\_color $((color_z, \mathcal{T}_{color_z}), \lambda_j, id_j)$}

\begin{algorithmic}
\State \LineNo{07} \textbf{if} $(color_z, \lambda_j, id_j) \notin Candidat\_list_i$ \textbf{then}
\State \LineNo{08} \quad add$(\mathcal{T}_{color_z}, (color_z, \lambda_j, id_j), Candidat\_list_i)$;
\State \LineNo{09} \textbf{endif}
\State \LineNo{10} \textbf{if} $(\exists(color_z, \lambda_k, id_k) \in Candidat\_list_i) \wedge ((color_z, \lambda_j, id_j) > (color_z, \lambda_k, id_k))$ \textbf{then}
\State \LineNo{11} \quad sub$(\mathcal{T}_{color_z}, (color_z, \lambda_k, id_k), (color_z, \lambda_j, id_j), Candidat\_list_i)$;
\State \LineNo{12} \textbf{endif}
\State \LineNo{13} $\mathcal{CC}$broadcast first\_confirm$(Candidat\_list_i)$;
\State \LineNo{14} $Candidat\_list_i = \emptyset$;
\end{algorithmic}

\vspace{0.2cm}
\noindent\textbf{At the reception of first\_confirm $(Candidat\_list_j)$}

\begin{algorithmic}
\State \LineNo{15} \textbf{for each} $(color, \lambda, id) \in Candidat\_list_j$
\State \LineNo{16} \quad \textbf{if} $(id = id_i)$ \textbf{then}
\State \LineNo{17} \qquad $cnt_{color} = cnt_{color} + 1$;
\State \LineNo{18} \qquad \textbf{if} $(cnt_{color} = \Delta_i)$ \textbf{then}
\State \LineNo{19} \qquad \quad add$((color_j, \mathcal{T}_{color_j}), Bcolors_i)$;
\State \LineNo{20} \qquad \textbf{endif}
\State \LineNo{21} \quad \textbf{endif}
\State \LineNo{22} \textbf{endfor}
\State \LineNo{23} $\mathcal{CC}$broadcast last\_confirm$(Bcolors_i, 1)$;
\end{algorithmic}

\vspace{0.2cm}
\noindent\textbf{At the reception of last\_confirm $(Bcolors_j, hop)$}

\begin{algorithmic}
\State \LineNo{24} \textbf{if} $(hop = 1)$ \textbf{then}
\State \LineNo{25} \quad \textbf{for each} $(color, \mathcal{T}) \in Bcolors_i$
\State \LineNo{26} \qquad add$(\mathcal{T}, color, d1colors_i)$;
\State \LineNo{27} \quad \textbf{endfor}
\State \LineNo{28} \quad $\mathcal{CC}$broadcast last\_confirm$(Bcolors_j, hop+1)$;
\State \LineNo{29} \textbf{endif}
\State \LineNo{30} \textbf{if} $(hop = 2)$ \textbf{then}
\State \LineNo{31} \quad \textbf{for each} $(color, \mathcal{T}) \in Bcolors_i$
\State \LineNo{32} \qquad add$(\mathcal{T}, color, d2colors_i)$;
\State \LineNo{33} \quad \textbf{endfor}
\State \LineNo{34} \textbf{endif}
\end{algorithmic}

\end{algorithm*}

\subsection{Definitions and local variables}
For clarity, we provide the following definitions of terms used in the algorithm:

\begin{itemize}
   \vspace{4pt}
    \item $\mathcal{CC}$broadcast is a collision and conflict free broadcast operation performed by node $p_i$. This operation is only possible if $\mathit{Cl}$ \textbf{mod} $\chi = color_i$.
   \vspace{4pt}
    \item $\mathcal{T}\_\mathcal{CC}$broadcast is a $\mathcal{CC}$broadcast, this operation is possible only if $\mathit{Cl} < \mathcal{T}$.
       \vspace{4pt}
      \item  {\sc add$(time, set1, set2)$} is the operation of adding $set1$ into $set2$ at the current time and delete $set1$ from $set2$ at time $time$.
         \vspace{4pt}
      \item  {\sc sub$(time, set1, set2, set3)$} is the operation of substituting $set1$ by $set2$ in $set3$. This operation is possible only if $\mathit{Cl} < time$.
       \vspace{4pt}
      \item We say that the tuple $(a, b_1, c_1)$ has high priority than the tuple $(a, b_2, c_2, )$ and we write $(a, b_1, c_1) \succ (a, b_2, c_2)$ if $b_1 > b_2 $.
         \vspace{4pt}
\end{itemize}

For a given node $p_i$, we associate local variables. These are defined as follows:

\begin{itemize}
    \item $color_i$ denotes the permanent color assigned to node $p_i$ as determined by the algorithm presented in \cite{L3}.
       \vspace{4pt}
    \item $\mathcal{T}_{color}$ represents the expiration time at which all borrowers of the color $color$ must cease using it.
       \vspace{4pt}
    \item $Candidat\_list_i$ is a list composed of 3-tuples, each of the form $(color, \lambda, id)$, where:
    \begin{itemize}
        \item $color$ is a color currently lent by a neighboring node identified by $id$,
        \item $\lambda$ is the rate (or weight) associated with the node $id$.
    \end{itemize}
       \vspace{4pt}
    \item $Bcolor_i$ is a list of 2-tuples representing temporarily borrowed colors. Each tuple has the form $(color, \mathcal{T}_{color})$, where $\mathcal{T}_{color}$ indicates the expiration time of the borrowing.
       \vspace{4pt}
    \item $d1colors_i$ is the set of colors used by the first-hop neighbors of node $i$, denoted as $neighbors_i^1$, while $d2colors_i$ is the set of colors used by the second-hop neighbors, denoted as $neighbors_i^2$.
       \vspace{4pt}
    \item Each node maintains a counter $cnt_{color}$ for every color $color \in \{0, 1, \ldots, K\}$, which tracks relevant statistics or usage pertaining to that color.
\end{itemize}

\subsection{Description of the algorithm}

The node that decides to lend its color executes the procedure {\sc color\_lending()}. In this procedure, the lender node $p_i$, when it is its turn to broadcast, checks if there are \textit{eligible neighbors} (i.e., $Ncl^{color_i}_i \ne \emptyset$) that can take its color in the next rounds (line 01). If $p_i$ has eligible neighbors, it sends the message {\sc Lend\_color($(color_i, \mathcal{T}_{color_i}), Ncl^{color_i}_i$)} to those eligible neighbors. This message uses the composed argument $(color_i, \mathcal{T}_{color_i})$, which contains the color and when the borrowers of that color must stop using it. The second argument is the list containing the IDs of the eligible neighbors. Then the node sleeps to save energy (line 02).The procedure {\sc adding\_lend\_color()} is executed by nodes that receive the messages {\sc Lend\_color()}, {\sc interest\_in\_color()}, {\sc first\_confirm()} or the message {\sc last\_confirm()}. 

When receiving the message {\sc Lend\_color($(color_j, \mathcal{T}_{color_j}), Ncl^{color_j}_j$)}, the node $p_i$ ignores the message if $id_i$ is not in the list $Ncl^{color_j}_j$ (line 05). If $id_i$ is in the list $Ncl^{color_j}_j$, the node $p_i$ sends the message {\sc interest\_in\_color} $((color_z, \mathcal{T}_{color_z}), \lambda_i, id_i)$ (line 06). This message can be interpreted as a request for permission from its neighbors to take $color_z$.

Upon receiving the message {\sc interest\_in\_color} $((color_z, \mathcal{T}_{color_z}), \lambda_j, id_j)$, the node adds the $id$ of the message sender to the list of candidates $Candidat\_list_i$ if $id$ is the first candidate for $color_z$ (lines 07,08). Then node $p_i$ chooses the best candidate for $color_z$ (line 10). Then it sends the list of best candidates for each color (line 13). The best candidate for a color $color_z$ is the candidate node that broadcasts the maximum number of packets (i.e., the highest $\lambda$). In line 14, the node clears the list $Candidate\_list_i$. 

After receiving {\sc first\_confirm($Candidat\_list_j$)}, each node in $Candidat\_list_j$ checks if it has been chosen as the best candidate by all its neighbors. If so, it adds the borrowed colors to $Bcolors_i$ (lines 17-19). Finally, it sends the message {\sc last\_confirm($Bcolors_j, 1$)} to inform its neighbors and the neighbors of neighbors about its temporary colors (line 22). Upon receiving the message {\sc last\_confirm($Bcolors_j, hop$)}, node $p_i$ updates its list $d1color_i$ if $hop=1$. If $hop=2$, node $p_i$ updates its list $d2colors_i$ and does nothing else (lines 24-30).

\subsection{Determining  the color lending duration}

Here, we discuss a suitable value for \(\mathcal{T}_{\text{color}_i}\) for a lender node \(p_i\), which ensures that it does not miss newly detected packets with high probability. Note that if node \(p_i\) permanently leaves the network, then we set \(\mathcal{T}_{\text{color}_i} \to +\infty\), allowing the borrowers to use the color indefinitely.

Recall that node \(p_i\) captures packets according to a Poisson process of rate \(\lambda_i\). If the node detects more than \(M\) packets before the round satisfying the condition \(\mathit{Cl} \bmod \chi = \text{color}_i\), the excess packets are lost.

Let \(X_{1,i}, X_{2,i}, \ldots\) be the interarrival times of sensed packets for node \(p_i\), where \(X_{m,i}\) is the time elapsed between the \((m-1)\)-th and the \(m\)-th sensed packet. Define \\
\begin{equation}
S_{m,i} \;=\; X_{1,i} + X_{2,i} + \cdots + X_{m,i}
\end{equation} \\
as the waiting time until the \(m\)-th packet is sensed by node \(p_i\). Then \(S_{m,i}\) follows a \(\mathrm{Gamma}(m,\lambda_i)\) distribution, whose mean is
\begin{equation}
\underbrace{\frac{1}{\lambda_i} + \cdots + \frac{1}{\lambda_i}}_{m\text{ times}}
\;=\; \frac{m}{\lambda_i}. \\
\end{equation} \\
Hence, if node \(p_i\) uses a single slot of duration \(\chi\) for transmission, the expected time remaining for the \(m\)-th captured packet to be transmitted is
\begin{equation}
\chi \;-\; \frac{m}{\lambda_i}.
\end{equation}

Next, define \(N_i(t)\) as the largest integer \(m\) such that \(S_{m,i} \le t\). The number of arrival events in any interval of length \(t\) satisfies:
\begin{equation}
\mathbb{P}\bigl(N_i(t+s)-N_i(s) = k\bigr)
\;=\;
\frac{(\lambda_i\,t)^k}{k!} \, e^{-\lambda_i\,t}, 
\quad k = 0,1,2,\ldots
\end{equation}

Each time node \(p_i\) transmits its buffered packets, the buffer is flushed and the process regenerates. Because the interarrival times \(X_{1,i}, X_{2,i}, \ldots\) are exponentially distributed with parameter \(\lambda_i\), the probability that no packets arrive for node \(p_i\) in the time interval \([\mathit{Cl},\,\mathcal{T}_{\text{color}_i}]\) is
\begin{equation}
\exp\bigl(-\lambda_i\,(\mathcal{T}_{\text{color}_i} - \mathit{Cl})\bigr).
\end{equation}
More generally, the probability of exactly \(x\) arrivals in an interval of length \(t\) is
\begin{equation}
\mathbb{P}\bigl(N_i(t) = x\bigr)
\;=\;
\frac{(\lambda_i\,t)^x}{x!}\,e^{-\lambda_i\,t}.
\end{equation}

Therefore, the probability that no packets arrive in \([\mathit{Cl},\,\mathcal{T}_{\text{color}_i}]\) is
\begin{equation}
\mathbb{P}\bigl(N_i(\mathcal{T}_{\text{color}_i}) - N_i(\mathit{Cl}) = 0\bigr)
\;=\;
\exp\bigl(-\lambda_i\,(\mathcal{T}_{\text{color}_i} - \mathit{Cl})\bigr).
\end{equation}

For node \(p_i\) and a user-defined parameter \(\alpha>0\), requiring that the probability of losing any packet is at most \(1/\Delta^\alpha\) implies
\begin{equation}
\exp\bigl(-\lambda_i\,(\mathit{Cl} - \mathcal{T}_{\text{color}_i})\bigr)
\;=\;
\frac{1}{\Delta^\alpha}.
\end{equation}
Taking the natural logarithm yields
\begin{equation}
\mathit{Cl} - \mathcal{T}_{\text{color}_i}
\;=\;
\frac{\alpha\,\ln(\Delta)}{\lambda_i}.
\end{equation}
This choice guarantees that the probability of any packet loss during the lending period remains bounded by \(1/\Delta^\alpha\). The parameter \(\alpha\) controls the trade‐off between reliability (packet loss probability \(1/\Delta^\alpha\)) and lending duration: the larger \(\alpha\) is, the shorter the lending period and the lower the loss probability.

\section{Experimental Results}
In this section, simulations are performed to verify the effectiveness of the proposed protocol. We consider a $1000 \ m^2$ area in which a random deployment of wireless nodes with a neighborhood of $200 m$ takes place. At the beginning, each node is assigned a unique time slot within its 2-hop neighborhood. The duration of each round is fixed such that $\chi t_{round} = 1 ms$, where $\chi$ represents the number of unique time slots. We categorize the nodes into two types: nodes with high sampling frequency (i.e., high values of $\lambda$, in the range of 4000-5000 times per minute) and nodes with low sampling frequency (i.e., low values of $\lambda$, in the range of 10-50 times per minute).

In the proposed protocol, only nodes with low values of $\lambda$ are allowed to lend their time slots. Each lending node calculates the lending time. A lending node lends its time slot to a set of eligible (non-collision causing) nodes from its neighborhood and sleeps for that period of time. After waking up, the node waits until it receives sensing information to perform the next lending operation. All nodes have the same memory size $m_{size} = 5$, which can store up to 5 packets at a time. Each node $p_i$ receives sensing packets according to a Poisson process with rate $\lambda_i$. If the memory is full at the time of reception, the node will drop the sensed packet due to a memory overflow. When the time slot occupied by the node arrives, the latter is able to transmit all the packets stored in its memory at that time slot.

We study the impact of our algorithm on the network performance compared to the normal operation of the network in terms of packet loss, packet waiting time (i.e., before broadcasting), and energy gain (resulting from sleep time). We also analyze the performance of our algorithm under different configurations.

\subsection{Packet loss}
\begin{figure}[t]
        \centering
        \includegraphics[width=0.42\textwidth]{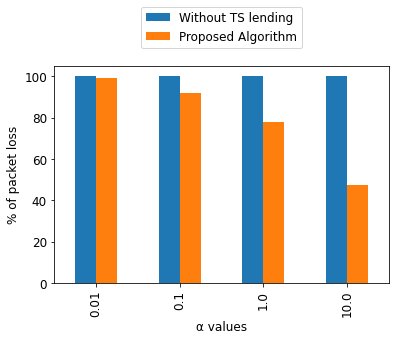}
        \caption{Average packet loss for nodes with more than one allotted time slot for different values of $\alpha$.}
        \label{fig1}
    \end{figure}

First, we show the impact of our proposed protocol on packet loss due to memory overflow. Figure \ref{fig1} depicts the percentage of packet loss for nodes with more than one allocated time slot compared to normal operation (i.e., a unique slot).

For different values of $\alpha$, the average percentage of packet loss at the level of these nodes decreases as $\alpha$ increases. This is due to the increase in allocation time at the level of the nodes that lend the time slots, allowing the nodes with different time slots to make additional broadcasts before memory overflow. For $\alpha$ values of 0.1, 1, and 10, packet loss can be reduced by more than 10\%, 20\%, and 50\% at the level of the borrowing nodes. 

According to the above results, we can conclude that our algorithm can improve the robustness and reliability of the network by reducing the packet loss.

\subsection{Packet waiting time}

\begin{figure}[t]
        \centering
        \includegraphics[width=0.42\textwidth]{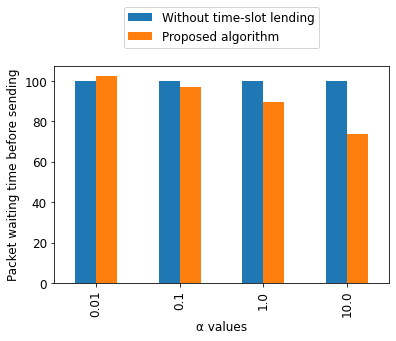}
        \caption{Packet waiting time before sending for different values of $\alpha$.}
        \label{fig3}
    \end{figure}

Upon receiving a sensed packet, the node temporarily stores it in its memory until the designated time slot for a broadcast arrives. The efficiency of the network is directly affected by the duration of this waiting period. To evaluate efficiency, we calculate and compare the time interval between packet arrival and packet broadcast in a network implementing our solution compared to a conventional network. The results are shown in Figure \ref{fig3}. The results show that, on average, the packet waiting time at the level of nodes with borrowed slots is reduced by more than 5\%, 10\%, and 20\%, respectively, for $\alpha$ values of 0.1, 1, and 10. 

This reduction means that our algorithm effectively reduces the waiting time of packets before transmission, which is crucial for real-time applications. By minimizing waiting time, our algorithm improves overall network performance and responsiveness, especially for time-sensitive applications that require prompt data delivery.

\subsection{Energy gain}
\begin{figure}[t]
        \centering
        \includegraphics[width=0.45\textwidth]{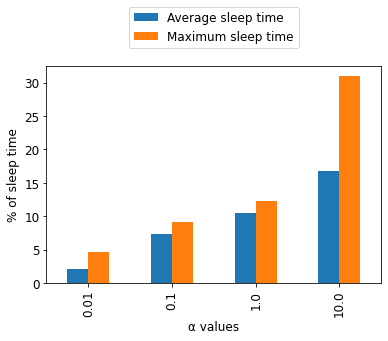}
        \caption{Percentage of average and maximum sleep time of nodes for different values of $\alpha$.}
        \label{fig4}
    \end{figure}

   In order to optimize energy consumption, nodes have the ability to sleep while leasing their time slots to other nodes. The percentage of time spent sleeping directly affects the amount of energy that can be saved. Figure \ref{fig4} shows the average and maximum percentage of sleep time for nodes that lend their time slots in the network, considering different values of $\alpha$. The results show that nodes can sleep for up to 30\% of the time, resulting in significant energy savings. 
   
   By allowing nodes to sleep during the lending process, our algorithm promotes energy efficiency in the network. This feature is particularly beneficial in resource-constrained scenarios, as it extends the lifetime of individual nodes and improves the overall sustainability of the network.
    
\section{Conclusion}
As constrained wireless networks continue to grow in scale and complexity, designing efficient and reliable channel access protocols becomes increasingly challenging, particularly under stringent QoS and energy constraints. In this paper, we proposed a novel slot-lending mechanism that dynamically reclaims underutilized time slots and reallocates them to active nodes. Our approach includes an efficient time slot allocation algorithm that improves both channel utilization and communication reliability. Experimental results demonstrate the effectiveness of the proposed solution, confirming its potential to enhance overall network performance in resource-constrained environments.

For future work, we plan to conduct a comprehensive robustness analysis of the proposed protocol under various network conditions. In particular, we aim to evaluate its resilience in scenarios involving node failures, mobility, fluctuating network densities, and increased interference levels. These evaluations will provide deeper insights into the adaptability of our approach. Additionally, we intend to explore the integration of Machine Learning (ML) techniques to further optimize slot lending decisions, potentially replacing or augmenting the current probabilistic mechanism with more context-aware and data-driven strategies.

\balance

\end{document}